\documentclass[a4paper,12pt]{article}
\usepackage{epsfig,float}
\usepackage[left=2cm,right=2cm]{geometry}
\usepackage{a4wide,graphicx}

\newcommand{\delmass}{M_{\Delta}} 

\newcommand{\nn}{\nonumber}

\newcommand{\be}{\begin{equation}}
\newcommand{\ee}{\end{equation}}
\newcommand{\ba}{\begin{eqnarray}}
\newcommand{\ea}{\end{eqnarray}}

\newcommand{\ov}{\overline}

\newcommand{\De}{\Delta}

\newcommand{\Sg}{\Sigma^*}

\newcommand{\X}{\Xi^*}

\newcommand{\Om}{\Omega}

\newcommand{\ksb}{\ov {K^*}}
\newcommand{\ep}{\epsilon}

\begin{document}

\title{Dynamically generated resonances from the vector octet-baryon decuplet
interaction}

\author{Sourav Sarkar$^{1,2}$, Bao-Xi Sun$^{1,3}$, E. Oset$^1$ and M.J. Vicente
Vacas$^1$ }
\maketitle

\begin{center}
$^1$ Departamento de F\'{\i}sica Te\'orica and IFIC,
Centro Mixto Universidad de Valencia-CSIC,
Institutos de Investigaci\'on de Paterna, Aptdo. 22085, 46071 Valencia, Spain \\
$^2$ Variable Energy Cyclotron Centre, 1/AF, Bidhannagar, Kolkata 700064, India \\
$^3$ Institute of Theoretical Physics, College of Applied Sciences,\\
Beijing University of Technology, Beijing 100124, China
\end{center}

\date{}

 \begin{abstract}
We study the interaction of the octet of vector mesons with the decuplet of baryons using Lagrangians of the hidden gauge theory for vector interactions. The unitary amplitudes in coupled channels develop poles that can be associated with some known baryonic resonances, while there are predictions for new ones at the energy frontier of the  experimental research. The work offers guidelines on how to search for these resonances.
\end{abstract}

\section{Introduction}

 The combination of chiral Lagrangians with nonperturbative unitary techniques
in coupled channels has been very fruitful and has provided a powerful tool to
study meson meson and meson baryon interactions beyond the realm of
applicability of chiral perturbation theory. It leads to accurate cross
sections for meson meson and meson baryon interactions and allows one to
study the analytical properties of the scattering matrix, where poles are
sometimes found, which can be associated to known resonances or new ones. Examples
of this are the low lying scalar states found in the case of meson meson
interaction  \cite{npa,prd,Kaiser:1998fi,Markushin:2000fa}, the $J^P=1/2^-$
low lying baryon states found in the interaction of pseudoscalar mesons with
baryons of the octet of the $p$
\cite{Kaiser:1995cy,weise,Kaiser:1996js,angels,ollerulf,carmina,carmenjuan,hyodo,Hyodo:2006kg,inoue},
and the   $J^P=3/2^-$ states obtained form the interaction of pseudoscalar
mesons with the decuplet of baryons of the $\Delta$ \cite{kolodecu,sarkar}. As
an example of new states predicted, one has a second $\Lambda(1405)$ state
\cite{cola}, right now reconfirmed by all the chiral unitary calculations, and
which finds  experimental support from the analysis of the $K^- p \to \pi^0 \pi^0
\Sigma^0$ reaction \cite{Prakhov:2004an} done in \cite{magas}. It also finds
an explanation for the new COSY data on $\Lambda(1405)$ production in $pp$ collisions
\cite{Zychor:2007gf} as shown in \cite{Geng:2007hz}.

  The work with vector mesons was for some time limited to the study of their
  interaction with pseudoscalar mesons, which generates the low
  lying axial vector mesons \cite{Lutz:2003fm,Roca:2005nm} and a second $K_1(1270)$
  state \cite{Roca:2005nm}, for which experimental evidence is found in
  \cite{Geng:2006yb}.

     A qualitative step in the interaction of vector mesons and generation of
resonances was done in \cite{Molina:2008jw}, where by using the Lagrangians of the
hidden gauge approach to vector interactions \cite{hidden1,hidden2,hidden3}, one scalar
and a tensor meson
were found from the $\rho \rho$ interactions, which could be identified with
the $f_0(1370)$ and $f_2(1270)$ mesons, and
for which the partial decay width into the sensitive $\gamma \gamma$ decay channel
was found to agree with experiment \cite{Nagahiro:2008um}.  An extension to the full
interaction of the octet of vector mesons among themselves has been recently done
\cite{gengvector}, indicating that several scalar, axial vectors and tensor states
reported in the PDG \cite{Amsler:2008zz} qualify as dynamically generated states
from the vector-vector interaction.

     In the baryon sector the interaction of vector mesons with baryons leading to
generation of resonances was first done in \cite{vijande} for the case of the
 $\rho \Delta$ interaction, which leads to three $N^*$ and three $\Delta^*$ states
 in the vicinity of 1900 MeV, which are degenerate in  $J^P=1/2^-,3/2^-,5/2^-$
 within the approximations done, and which can be associated to existing states
 of the PDG.

    After the successful predictions for the lowest states generated with vector
mesons and baryons of the decuplet  with $\rho \Delta$, the extension to the full
$SU(3)$ space of vectors
and members of the decuplet of baryons is most opportune and this is the purpose of
the present work.  We extend the work of \cite{vijande}, using the formalism
developed there, and we find {\em ten} states dynamically generated, some of which
can be clearly associated to known resonances, while others are less clear. The
states found are also degenerate in spin,
and in some cases the known states of the PDG support this finding. In other cases
one or two of these spin states are found in the PDG but others are missing. The
findings of the present work indicate that such spin parters should exist and
this offers a new motivation for the search of new baryonic resonances with the
quantum numbers predicted.

\section{Formalism for $VV$ interaction}

We follow the formalism of the hidden gauge interaction for vector mesons of
\cite{hidden1,hidden2,hidden3} (see also \cite{hidekoroca} for a practical set of Feynman rules).
The Lagrangian involving the interaction of
vector mesons amongst themselves is given by
\begin{equation}
{\cal L}_{III}=-\frac{1}{4}\langle V_{\mu \nu}V^{\mu\nu}\rangle \ ,
\label{lVV}
\end{equation}
where the symbol $\langle \rangle$ stands for the trace in the $SU(3)$ space
and $V_{\mu\nu}$ is given by
\begin{equation}
V_{\mu\nu}=\partial_{\mu} V_\nu -\partial_\nu V_\mu -ig[V_\mu,V_\nu]\ ,
\label{Vmunu}
\end{equation}
where  $g$ is given by
\begin{equation}
g=\frac{M_V}{2f}\ ,
\label{g}
\end{equation}
with $f=93\,MeV$ the pion decay constant. $V_\mu$ corresponds to the $SU(3)$
matrix of the vectors of the octet of the $\rho$ with the mixing of $\phi$
and $\omega$ accounted for and is given by
\begin{equation}
V_\mu=\left(
\begin{array}{ccc}
\frac{\rho^0}{\sqrt{2}}+\frac{\omega}{\sqrt{2}}&\rho^+& K^{*+}\\
\rho^-& -\frac{\rho^0}{\sqrt{2}}+\frac{\omega}{\sqrt{2}}&K^{*0}\\
K^{*-}& \bar{K}^{*0}&\phi\\
\end{array}
\right)_\mu \ .
\label{Vmu}
\end{equation}

The interaction of ${\cal L}_{III}$ gives rise
to a three
vector vertex form
\begin{equation}
{\cal L}^{(3V)}_{III}=ig\langle (\partial_\mu V_\nu -\partial_\nu V_\mu) V^\mu V^\nu\rangle
\label{l3V}\ ,
\end{equation}
which can be conveniently rewritten using the property of the trace as

\begin{eqnarray}
{\cal L}^{(3V)}_{III}=ig\langle [V^\mu(\partial_\nu V_\mu) -(\partial_\nu V_\mu)
V^\mu] V^\nu\rangle
\label{l3Vsimp}\ .
\end{eqnarray}

\begin{figure}[htb]
\begin{center}
\includegraphics[width=0.7\textwidth]{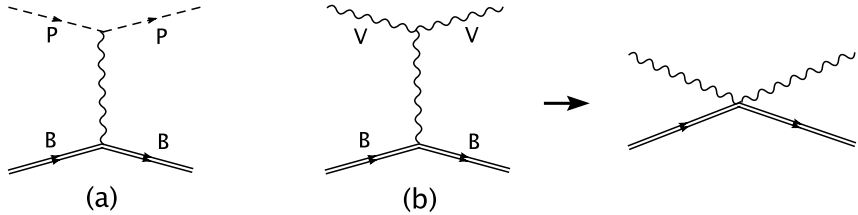}
\end{center}
\caption{Diagrams contributing to the pseudoscalar-baryon (a) or vector-
baryon (b) interaction via the exchange of a vector meson leading to the effective
vector-baryon contact interaction which is used in the Bethe-Salpeter equation}
\label{fig:feyn}
\end{figure}

With the three vector coupling we can construct the Feynman diagram which is
responsible for the vector baryon decuplet interaction through the exchange of
a vector between two vectors and the baryon, see fig.~\ref{fig:feyn}(b),
in analogy with the interaction of a pseudoscalar meson with the baryon decuplet
depicted in fig.~\ref{fig:feyn}(a). This mechanism of interaction of
pseudoscalars with baryons, leads
to the well known Weinberg-Tomozawa term, upon neglecting
$q^2/M_V^2$ in the propagator of the exchanged vector, where $q$ is the
momentum transfer. The local chiral Lagrangian for vector baryon decuplet interaction
 can thus be obtained from this
formalism upon neglecting the momentum transfer compared to the mass of the vector,
a good approximation for the purpose of studying the interaction
relatively close to threshold where bound states or resonances are searched
for. Consistent with this implicit approximation of the chiral Lagrangians,
we neglect the three-momentum of the vectors compared to their mass.  In this
approximation the polarization vectors of the vector mesons have only spatial
components. This simplifies considerably the formalism. Indeed, let us look at
eq. (\ref{l3Vsimp}) where we see that the field $V^\nu$   cannot
correspond to an external vector meson. Indeed, if this were the case, the $\nu$
index should be spatial and then the partial derivative $\partial_\nu$
would lead to
a three momentum of the vector mesons which are neglected in the approach. Then
 $V^\nu$ corresponds to the exchanged vector. In this case one finds an analogy to the coupling of vectors to pseudoscalars given in the same theory by

 \begin{center}
\be
{\cal L}_{VPP}=-ig\langle [P(\partial_\nu P) -(\partial_\nu P)P] V^\nu\rangle,
\label{lagrVpp}
\ee
\end{center}
where $P$ is the $SU(3)$ matrix of the pseudoscalar fields. The only difference
is the polarization vector of the two external fields and a sign, which
recalling that we only have spatial components for the external vectors,
results in the factor $\vec{\epsilon}\cdot\vec{\epsilon '}$ additional to the
contribution from the $PPV$ Lagrangian of eq. (\ref {lagrVpp}), from the upper
vertex of fig. 1 (a).

  Consequently, in order to obtain the tree level amplitudes corresponding to
the diagram of fig. 1(b), all one has to do is to take the corresponding
amplitudes of pseudoscalar meson-baryon decuplet interaction, substituting the
$SU(3)$ pseudoscalar matrix by the corresponding one of the vector mesons. That
is, $\pi^+ \Delta^+ \to \pi^+ \Delta^+$ is  substituted by
$\rho^+ \Delta^+ \to \rho^+ \Delta^+$ and so on.

   A small amendment is in order, which
   is due to the mixing of $\omega_8$ and the singlet of $SU(3)$, $\omega_1$, to give the
   physical states of $\omega$ and $\phi$

   \begin{eqnarray}
   \omega=\frac{2}{\sqrt 6} \omega_1 + \frac{1}{\sqrt 3} \omega_8 \nonumber \\
   \phi=\frac{1}{\sqrt 3} \omega_1 - \frac{2}{\sqrt 6} \omega_8
   \label{eq:omephi}
   \end{eqnarray}
   Given the structure of eq.~(\ref{eq:omephi}), the singlet state  which is accounted for
   by the $V$ matrix, $diag(\omega_1,\omega_1,\omega_1)/\sqrt3$, does not provide
   any contribution to eq.~(\ref{l3Vsimp}), in which case, one simply has to take the
   matrix elements known for the $PB$ interaction and wherever $P$ is the
   $\eta_8$ multiply the amplitude by the factor $1/\sqrt 3$ to get the
   corresponding $\omega $ contribution and by $-\sqrt{ 2/3}$ to get the
   corresponding $\phi$ contribution.

      We take the matrix elements for the amplitudes from~\cite{sarkar}.
There the same approximations that we make for the vector mesons, neglecting
the the three-momentum versus their mass, were also done for the baryons and
then all the amplitudes have the form

   \begin{equation}
V_{i j}= - C_{i j} \, \frac{1}{4 f^2} \, (k^0 + k'^0)~ \vec{\epsilon} \cdot \vec{\epsilon
^\prime} ,
\label{eq:kernel}
\end{equation}
where $k^0, k'^0$ are the energies of the incoming and outgoing vector meson
respectively. The amplitudes are thus exactly the ones for $PB\to PB$
apart for the factor $  \vec{\epsilon} \cdot \vec{\epsilon'}$.

    The $C_{ij}$ coefficients of eq. (\ref{eq:kernel}) can be obtained directly from
    \cite{sarkar}
    with the simple rules given above for the $\omega$ and the $\phi$, and
    substituting $\pi$ by $\rho$ and $K$ by $K^*$ in the matrix elements. The
    coefficients are obtained both in the physical basis of states and in the
    isospin basis. Here we will directly study the interaction in isospin
    basis and we collect the tables of the $C_{ij}$ coefficients in the
    Appendix for different states of isospin, $I$, and strangeness, $S$.

     Inspection of the tables immediately show the cases where we find
     attractive interaction and we can expect bound states or resonances, and where
 there is repulsion no states should be expected. We find attraction in the
 channels $S,I=0,1/2$; $0,3/2$; $-1,0$; $-1,1$; $-2,1/2$ and maybe $-3,0$,
  where the diagonal
 terms are zero but one can get bound states through the interplay of the
 non diagonal terms of the coupled channels. On the other hand in the $0,5/2$; $-1,2$;
 $-2,3/2$; $-3,1$ and $-4,1/2$ one finds repulsion and one should not expect bound states or resonances there.
 Interestingly, these are all exotic channels and we find that even if these
 exotic quantum numbers can be reached within the approach, for dynamical
 reasons, no bound states
 are found in either of the possible exotic channels.

\begin{figure}[htb]
\begin{center}
\includegraphics[width=0.7\textwidth]{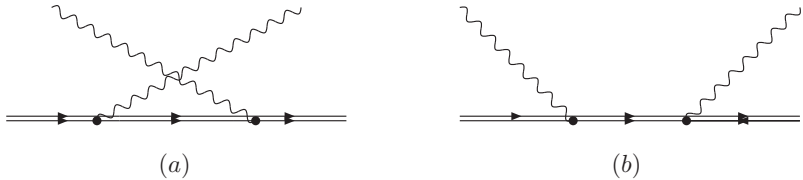}
\end{center}
\caption{(a) u-channel mechanism. (b) s-channel mechanism.}
\label{fig:u-s-channel}
\end{figure}

As one can see, we have only considered t-channel mechanisms with
the exchange of vector mesons. The strength of these terms is very
large, since it is roughly proportional to the sum of two vector
masses, by comparison to the interaction of pseudoscalar mesons with
baryons, where the interaction is proportional to the sum of two
pseudoscalar masses. One may wonder what happens with other
mechanisms of u-channel type. The u-channel mechanism with the
building blocks that we are using is depicted in
fig.~\ref{fig:u-s-channel}$(a)$. Similarly we could also consider
the s-channel depicted in Fig.~\ref{fig:u-s-channel}$(b)$.
In both cases, one finds terms of order $q/M_V$ contributing mostly
to p-waves. Since we only consider s-wave interaction and
furthermore we neglect terms of $q/M_V$ in our approach, all these
terms are omitted for consistency of the whole approach. The part coming from
the negative energy components of the baryon propagators in 
fig.~\ref{fig:u-s-channel}$(b)$ contributes to the s-wave, but should also be
neglected for consistency with the nonrelativistic approach followed in the
interaction vertices. This contribution, evaluated in \cite{cabrera} (see
appendix of that paper),
is anyway small compared to the leading
terms that we have here. In the
s-wave mechanism one can think of the possibility of having negative
parity resonances in the intermediate state to which $\rho \Delta$
couple strongly, but these are the resonances that we are generating
dynamically and, thus, those terms should be omitted to avoid double
counting. We will come back to these terms in subsection 2.2.

    The next step to construct the scattering matrix is done by solving the
    coupled channels Bethe Salpeter equation in the on shell factorization approach of
    \cite{angels,ollerulf}
   \begin{equation}
T = [1 - V \, G]^{-1}\, V
\label{eq:Bethe}
\end{equation}
with $G$ the loop function of a vector meson and a baryon which we calculate in
dimensional regularization using the formula of \cite{ollerulf},
with $\mu$ a regularization scale of 700 MeV, and natural
values of the subtraction constants $a_l(\mu)$ around -2, as determined in
\cite{ollerulf}.

The on shell factorization has its root on the use of unitarity and
dispersion relations. It is an exact result if one omits the
contribution of the left hand cut, which is small for these
processes, and in any case smoothly energy dependent in the region
of interest to us, such that its contribution can be easily
accommodated in terms of a subtraction constant in the dispersion
integral~\cite{ollerulf} (details can be seen in this reference and
for a pedagogical overview in~\cite{pramana}, section 3). We will come back to
this point in subsection 2.2.

 The iteration of diagrams implicit in the Bethe Salpeter equation in the case
 of the vector mesons has a subtlety with respect to the case of the
 pseudoscalars. The $\vec{\epsilon} \cdot \vec{\epsilon^\prime}$ term of the interaction
 forces the intermediate vector mesons in the loops to propagate with the
 spatial components in the loops.  We need to sum over the polarizations of the
 internal vector mesons which, because they are tied to the external ones
 through the $\vec{\epsilon} \cdot \vec{\epsilon^\prime}$ factor, provides

 \begin{equation}
\sum_{pol} \epsilon_i \epsilon_j =\delta_{ij} + \frac{q_i q_j}{M_V^2}
 \end{equation}

As shown in \cite{Roca:2005nm}, the on shell factorization leads to a correction
coefficient in the $G$ function of ${\vec{q}}^{~2}/3 M_V^2$ versus unity, which is negligibly small,
and is also neglected here for consistency with the approximations done. In this
case the factor $\vec{\epsilon} \cdot \vec{\epsilon^\prime}$ appearing in the potential $V$,
 factorizes also in the $T$ matrix for the external vector mesons.

    Since the spin dependence only comes from the
  $\vec{\epsilon}\cdot\vec{\epsilon }'$ factor and there is no dependence on the spin
  of the baryons, the interaction for vector-baryon states with $1/2^-$, $3/2^-$
  and $5/2^-$ is the same and thus we get three degenerate states for each of
  the resonances found. The spin degeneracy also appears in some quark models
 \cite{kirchbach}.

  Finally, as done in \cite{vijande}, we take into account the convolution of
the $G$ function with the mass distributions of the $\rho$, $K^*$, $\Delta$,
$\Sigma(1385)$ and $\Xi(1530)$ states, in order to account for their sizable
width.

\subsection{Long range forces with intermediate vector nonet and baryon octet}

\begin{figure}[htb]
\begin{center}
\includegraphics[width=0.4\textwidth]{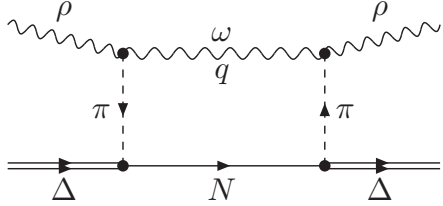}
\end{center}
\caption{Term with intermediate $\omega N$ in the $\rho \Delta
\rightarrow \rho \Delta$ interaction, involving the anomalous $\rho
\omega \pi$ coupling and pion exchange. } \label{fig:anomal}
\end{figure}

In Ref.~\cite{vijande}, in the study of the $\rho \Delta$
interaction, we evaluated explicitly the contribution of the box
diagram with an intermediate $\omega \Delta$ state, with the
transition $\rho \Delta \rightarrow \omega \Delta$ mediated by a
pion exchange. This term turned out to be small compared with the
contribution from $\rho$ exchange. The same long range force can
also produce $\omega N$ intermediate states, and in its extension to
$SU(3)$, states of a nonet vector meson and an octet baryon. We
evaluate explicitly here the case of $\rho \Delta \rightarrow \omega
N$, which involves the pion exchange. We will find a very small
contribution, and in other cases where we exchange heavier
pseudoscalar mesons one, thus, expects an even smaller contribution.
The diagram that we evaluate is shown in fig.~\ref{fig:anomal}.

We rely upon the evaluation of Ref.~\cite{vijande} and note here the
differences. The anomalous $\rho \omega \pi$ term is the same in
both cases. The only difference is the $\pi N \Delta$ vertex here
versus the $\pi \Delta \Delta$ used in Ref.~\cite{vijande}. The $\pi
N \Delta$ vertex is given by
\begin{equation}
-i t_{\pi N \Delta}~=~\frac{f_{\pi N \Delta}}{m_\pi}~\vec{S} \cdot
\vec{q}~\vec{T} \cdot \vec{\phi},
\end{equation}
where $\vec{S}$ and $\vec{T}$ are the spin and isospin transition
operators from $3/2$ to $1/2$, normalized such that
\begin{equation}
\langle 3/2, M |S^\dagger_\nu|1/2, m\rangle~=~C(1/2, 1, 3/2; m, \nu,
 M),
\end{equation}
and the same for $T^\dagger_\nu$, with $S^\dagger_\nu$ written in
spherical basis and $C(1/2, 1, 3/2; m, \nu, M)$ the ordinary
Clebsch-Gordan coefficients. 

We find easily that
\begin{equation}
\langle \Delta \rho, I=1/2 | \vec{T}^\dagger \cdot
\vec{\phi}~\vec{T} \cdot \vec{\phi} | \Delta \rho, I=1/2
\rangle~=~2.
\end{equation}
On the other hand, for the spin part, instead of $\vec{S}^2=3/2\cdot
5/2$ in Ref.~\cite{vijande}, we find now $\vec{S}^\dagger \cdot
\vec{S}=1$. Thus, we replace $$\frac{15}{4}\times \frac{15}{4}
f^2_\Delta \rightarrow 2\times 1 f_{\pi N \Delta}^2$$ with the value
$f_{\pi N \Delta}=2.23$ and $f_\Delta=0.802$. These two coefficients
are nearly equal. This is not the only change. In eq.~(37) of
Ref.~\cite{vijande}, one must in addition change $E_\Delta
\rightarrow E_N=\sqrt{\vec{q}^2+M_N^2}$, wherever it appears, caring
to put the $i\epsilon$ in the
$(P^0-k^0-\omega_\pi-E_N+i\epsilon)^{-2}$ term.

\begin{figure}[htb]
\begin{center}
\includegraphics[width=0.7\textwidth]{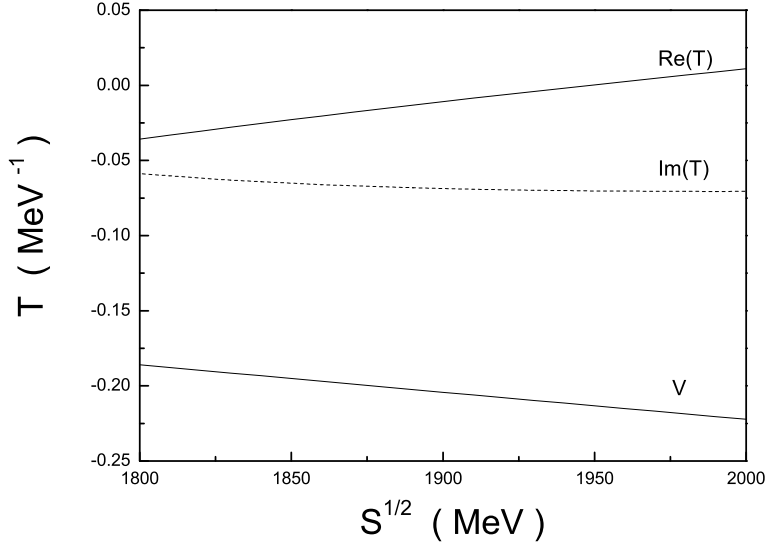}
\end{center}
\caption{ The real and imaginary parts of the anomalous $T_{\rho
\Delta \rightarrow \rho \Delta}$ with a intermediate $\omega N$
state compared to $V$ for $I=1/2$ as a function of $\sqrt{s}$ for
$q_{max}=770MeV$.} \label{fig:vv}
\end{figure}
The numerical results can be seen in fig.~\ref{fig:vv}. We can see
that compared to the term from the hidden gauge Lagrangian for
$I=1/2$ (eq.~(\ref{eq:kernel}) and Table \ref{tab:cijs0i12}), which is
shown in fig.~\ref{fig:vv}, the real part of the new term is very
small around $1850$ MeV where the $I=1/2$ appears, and we neglect 
it in the calculations. The imaginary part is
also small, but it would add to the width of the $I=1/2$ states.

\subsection{Considerations on u-channel and t-channel exchanges and the on-shell
factorization}

  The procedure followed here relies upon the on shell factorization of the
amplitudes for a given energy, assuming implicitly that they come from a contact
Lagrangian, and a coupled channel approach that implements unitarity in coupled
channels. The method uses coupled Bethe Salpeter equations, which in this 
situation can be converted into algebraic equations.  One way to deduce these
equations is to assume the N/D method implementing dispersion relations and
neglecting the contribution of the left hand cut \cite{nsd,ollerulf}. This is
justified at medium energies in the interaction of light 
particles with heavy ones, implicitly assuming that the mass of the light
particles is small compared to the mass of a hypothetical particle that is
exchanged in the t-channel to provide the interaction. This is well suited for
the scattering of pseudoscalar mesons off baryons \cite{ollerulf}. The fact that
we have here the scattering of vector mesons off baryons and that the
interaction is driven by the exchange of vector mesons indicates that we are not
in the ideal conditions like the scattering of pseudoscalar mesons and we have to
find a justification for the procedure followed.  To better understand the
problem let us go into it in more detail. 

   The leading term of the potential in our case comes from the mechanism
of  fig.~\ref{fig:feyn} where we have the explicit exchange of a vector meson.
Certainly, in the approximation where $q^2$ has been neglected in comparison with
$M_V^2$, with $q$ the momentum transfer between the incoming and scattered vectors,
 no singularities 
appear in the potential. However, removing this
approximation, for certain energies below threshold and extrapolating $q^2$ 
assuming $p^2=m^2$ for the external particles, one can have $q^2=M_V^2$ and the propagator develops a
singularity, giving rise to the branch point of the left hand cut associated with
the t-channel exchange. Similarly, if one takes the diagram of fig.
\ref{fig:u-s-channel}(a), and assuming again $p^2=m^2$ for the external legs,
the intermediate baryon can be placed on shell for a certain energy below
threshold. The hope is that in any case these left hand cuts are far away from
the energy region of interest and they lead to a weakly energy dependent 
contribution which can be accommodated by a subtraction constant in
the dispersion relation, the idea behind \cite{nsd,ollerulf}. This, however,  is
not the case here.

We begin with the t-exchange mechanism of 
fig. \ref{fig:feyn}. Let us take the case of $\rho \Delta\to\rho \Delta$ 
(we shall discuss
other channels later on). In this case, calling $k,k'$ the momenta of the
incoming, outgoing vectors and $q=k-k'$, we see that $q^0$  is zero, because it
is an elastic channel. On the other hand, we have with the on-shell prescription
of taking $p^2=m^2$ for the external particles,
\be
|\vec k|^2=\frac{[s-(m_\rho+M_\Delta)^2][s-(m_\rho-M_\Delta)^2]}{4s}
\label{eq:new1}
\ee
and the intermediate rho meson propagator develops a pole at the highest energy 
when $\vec{k}$ and $\vec{k}'$ have opposite sign and then
\be
-4|\vec k|^2-m_\rho^2=0
\label{eq:new2}
\ee
where $|\vec k|^2$ will be negative and $k$ purely imaginary, which leads to 
\be
s=\frac{1}{2}(2M_\Delta^2+m_\rho^2+\sqrt{12M_\Delta^2m_\rho^2-3m_\rho^4}~)
\label{eq:new3}~.
\ee
This gives us a branch point at $\sqrt s =1837~ $MeV.
We could consider this to be far away from the region of interest, except that
the lowest energy of the states that we get from this interaction has an 
energy around $1850~$ MeV. 

 Let us go deeper into the issue by really taking the on-shell
($p^2=m^2$) extrapolation below threshold and then projecting the $\rho$-exchange
potential over $s$-wave. This is done by means of 
\ba
V_s&=&\frac{1}{2}\int^1_{-1}d\cos\theta\frac{1}{-|\vec k|^2-|\vec k'|^2+2
|\vec k||\vec k'|\cos\theta-m_\rho^2}\nonumber\\
&=&\frac{1}{4|\vec k|^2}\ln\frac{m_\rho^2}{m_\rho^2+4|\vec k|^2}
\label{eq:new4}
\ea
with $|\vec k'|^2=|\vec k|^2$ given by eq.~(\ref{eq:new1}).  When we neglect $q^2$ 
this potential is
$(-m_{\rho}^2)^{-1}$ such that the ratio of the on-shell $s$-wave projected
potential to the static one is given by 
\be
R=\frac{m_\rho^2}{4|\vec k|^2}\ln\frac{m_\rho^2+4|\vec k|^2}{m_\rho^2}
\label{eq:new5}
\ee
eq.~(\ref{eq:new4}) shows the singularity which appears when eq.~(\ref{eq:new2}) 
is fulfilled. 
\begin{figure}[htb]
\begin{center}
\includegraphics[width=0.4\textwidth]{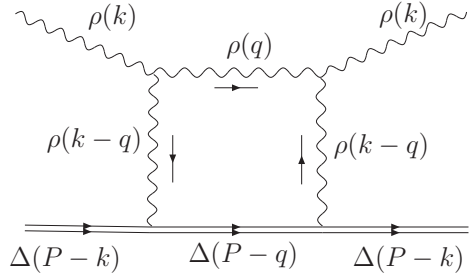}
\end{center}
\caption{Term with intermediate $\rho \De$ in the $\rho \Delta
\rightarrow \rho \Delta$ interaction, involving the $\rho\rho\rho$
coupling. } 
\label{fig:rho_loop}
\end{figure}
Once we reach this point it is interesting to visualize the meaning and extent
of the on-shell factorization of the N/D method. Let us now evaluate the 
Feynman box diagram of fig.~\ref{fig:rho_loop}. We take all vertices equal 
to unity since
we simply want to compare with the on-shell factorization. 
We evaluate the loop
function given by 
\ba
L&=&2iM_\Delta\int \frac{d^4q}{(2\pi)^4}\frac{1}{q^2-m_\rho^2+i\ep}
\left(\frac{1}{(k-q)^2-m_\rho^2+i\ep}\right)^2\frac{1}{(P-q)^2-M_\De^2+i\ep}\nn\\
&=&2iM_\Delta\frac{\partial}{\partial m_\rho^2}\int \frac{d^4q}{(2\pi)^4}
\frac{1}{q^2-m_\rho^{'\, 2}+i\ep}\frac{1}{(k-q)^2-m_\rho^2+i\ep}
\frac{1}{(P-q)^2-M_\De^2+i\ep}
\label{eq:new6}
\ea
where $m'_{\rho}$ will be made equal to $m_{\rho}$ after the derivative. Using
standard Feynman integral techniques we can write $L$ as 
\be
L=-\frac{1}{16\pi^2}2M_\De\int_0^1dx(1-x)\int_0^x\frac{dy}{(s'+i\ep)^2}
\label{eq:new7}
\ee
where
\be
s'=(x-1-x^2)m_\rho^2-sy^2+xy(s+m_\rho^2-M_\De^2)
\label{eq:new8}
\ee

\begin{figure}
\begin{center}
\includegraphics[width=0.6\textwidth]{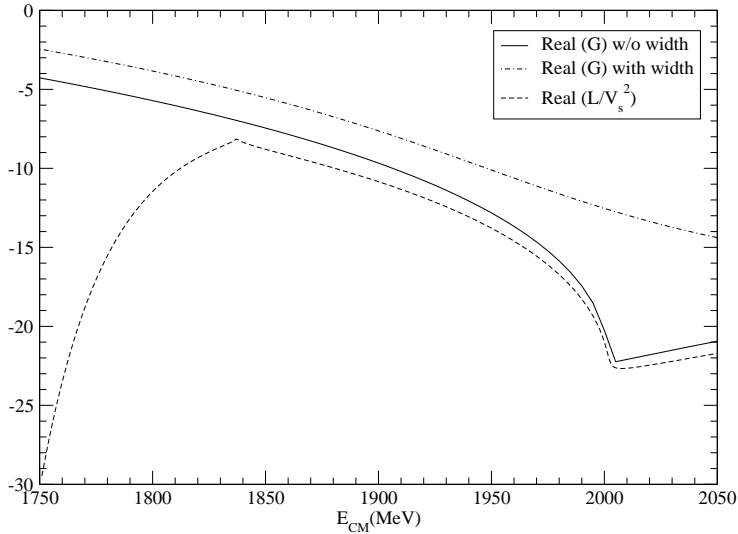}
\end{center}
\caption{$(L/V_s^2)$ as a function of the energy.  The $G$
 function after convolution with the spectral function of the $\rho$ and the
 $\Delta$ to account for their width is also shown for comparison.}
\label{fig:LvsG}
\end{figure} 
In the on-shell factorization this integral would correspond to 
\be
L\to V_s^2 \ G
\label{eq:new9}
\ee
where $G$ is the $\rho \Delta$ loop function of eq.~(\ref{eq:Bethe}).  
In fig.~\ref{fig:LvsG} we compare $L/V_s^2$ obtained from
eqs.~(\ref{eq:new4}) and (\ref{eq:new7}) with $G$ as a function of the
 energy. The idea
behind the on-shell factorization is that the two sides of eq.~(\ref{eq:new9})
 are
equivalent with a suitable subtraction constant in the function $G$, which
requires regularization since it is logarithmically divergent. 
 The subtraction constant
  in $G$ has
been chosen such that the two functions are equal at the $\rho \Delta$
threshold, the cusp point.
As we can
 see, the on-shell factorization and the Feynman integral are remarkably
 similar in the region of interest to us in this sector, 1850 MeV - 2000 MeV 
 \cite{vijande}, down to values of $s$ very close to the branch point
 of the t-channel. The differences are small compared to
 the effects of taking into account the  $\rho$ and the $\Delta$ mass
 distributions.  We will come back to this point again when mentioning coupled
 channels, where we shall recall the main point of the former discussion, which
 is the absolute failure of this on shell factorization below the branch point.

\begin{figure}
\begin{center}
\includegraphics[width=0.4\textwidth]{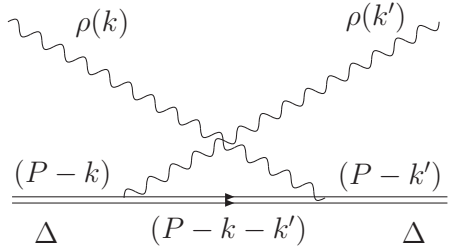}
\end{center}
\caption{$\Delta\rho$ scattering diagram in the u-channel.}
\label{fig:uchannel}
\end{figure}

 Let us now turn our attention to the left hand cut related to the u-channel.
 For this purpose let us look at fig.~\ref{fig:uchannel}. 
The intermediate $\Delta$ propagator reads 
 \be
 G_\De=\frac{1}{\sqrt{s}-\omega_\rho(k)-\omega_\rho(k')-E_\De(\vec k+\vec k')}
\label{eq:new10} 
 \ee
 Taking again the usual on-shell prescription of $p^2=m^2$ for the external legs
 we find the pole of
 eq.~(\ref{eq:new10}) at highest energy when $\vec{k}$ and $\vec{k'}$ are equal.
 Substituting the on-shell values of $k$ of eq.~(\ref{eq:new1}) we find the solution 
\be
s=\delmass^2+2m_\rho^2
\label{eq:new11}
\ee
 which sets the branch point of the u-channel at $\sqrt s = 1644~$MeV. This
 is far away from the region of interest so that we can safely neglect it. However,
 when considering coupled channels the branch point moves to higher
 energies in other channels. We will come back to this issue a little later.
 Should we have just
 this channel there is no cause for concern since we can incorporate
 s-wave contributions from this channel with a suitable subtraction constant in
 the $G$ function.  There is more to it since we can make explicit evaluations of
 the s-wave contribution of the u-channel. For this we project the propagator of
 eq.~(\ref{eq:new10}) over s-wave, as done in  eq.~(\ref{eq:new4}), taking the 
 $\gamma ^0 \epsilon ^0$ component, approximated by $k/M_V$ for the 
 $\rho \Delta $ coupling 
 (the spatial components lead to p-waves mostly)  and we find
 \be
 V_s(k)=\frac{\delmass}{2m_\rho^2}g^2\ln\frac{\sqrt s-2\omega_\rho(k)-\delmass}
 {\sqrt s-2\omega_\rho(k)-(\delmass^2+4|\vec k|^2)^{1/2}}
 \label{eq:new12}
 \ee
 If we compare this with the dominant t-channel $\rho$ exchange projected over
 s-wave, which is of the order of $2 g^2\omega_{\rho}(k) /(-m_{\rho}^2)$
 multiplied by $R$ of eq.~(\ref{eq:new5}), we see that the ratio is of the order 
 of 5 \% for $\sqrt{s}$
 above 1850  MeV and thus negligible. This is without counting extra $C$ factors
 of eq.~(\ref{eq:kernel}) , which tend to make the ratio even smaller.  For all
 these reasons the u-channel can be safely neglected when dealing with the
 s-wave interaction.  

    We now come  back to the t-channel branch points for the coupled channels.
 In the case of the $\rho \Delta$ interaction we have also the coupled channel
 $K^* \Sigma^*$.  In the diagonal transition $K^* \Sigma^* \to K^* \Sigma^*$ 
 we have also the $\rho$ as the lightest vector
 exchanged.  Substituting $\rho  \Delta$ by 
 $K^* \Sigma^*$ in eq.~(\ref{eq:new1}) together with eq.~(\ref{eq:new2})   
 lead now to  the branch point at $\sqrt s= 2133~$ MeV. This energy is 
 bigger than the energy of the two states that we obtain around 1850 MeV and
 1950 MeV \cite{vijande} and calls for some reflection. 
 As we could see from the discussion
 that lead to fig.~\ref{fig:LvsG}, the region of energies of interest to us
 around 1800 MeV-2000 MeV is about 250 MeV  below the branching point of the 
 $K^* \Sigma^*$ channel and the on shell factorization of the $\rho$ 
 exchange in the $K^* \Sigma^* \to K^* \Sigma^*$ potential would simply 
 be ruled out.
  This
 apparent failure is something which is welcome, because one can
 trace it to an abuse of the standard procedure used in one channel problems, 
 which needs to be changed for coupled channels. Indeed, the blind
 application of the on-shell factorization would imply using momenta of the
 $K^*$ of the order of $i300$ MeV; purely imaginary. Obviously this is
 meaningless from the physical point of view. The states that we have are a
 mixture of $\rho \Delta$ and $K^* \Sigma^*$ and have a certain energy, around
 1850 MeV and 1950 MeV \cite{vijande}, as we shall see. However, 
 the momentum of these
 components has to be a real quantity, which implies that the $K^* \Sigma^*$ are
 very bound. This is the physical world for bound states. The "on-shell" result
 that the $K^*$ has an imaginary momentum stems from the fact that we impose 
 $p^2=m^2$ for the $K^*$ and the  $\Sigma^*$, which is definitely not the case in
 the real situation. However, in one channel problems, the extrapolation to the 
 complex plane imposing  $p^2=m^2$ is a customary procedure. It is very useful
 because with the extrapolation to the complex plane one can make use of
  complex
 variable theorems and come out with valuable results from dispersion
 relations.  Yet, one should not forget that the extrapolated amplitudes 
 to the unphysical regions  have no physical meaning, and particularly the
 imaginary momenta cannot be confused with the actual momenta of the particles.
  
  We thus see that in the case of coupled channels, one must invoke the
  binding of the components, but the momenta will always be real. The size of
  the momenta will be typical of the momentum distribution of bound states of
  hadrons
  with masses similar to  those of the vector mesons. Nuclei are something of 
  this
  sort, with nucleons bound by about 60 MeV, and momenta  of the order
  of 200 MeV as average. If we take this range, $(\vec{q}/M_V)^2$ is of the
  order of 5 \% and a realistic "on-shell" factorization of the $\rho$
  propagator for the physical states would give essentially $-1/m_{\rho}^2$ .
  Obviously now $p^2\not=m^2$ for the external lines. The $K^*, \Sigma^*$ are
  now very bound, much like the $\Delta$ components are bound in ordinary nuclei
  \cite{Weber:1978dh,Moinester:1992qd}.
  
   There are more arguments to support the use of that approximation in the
   vector meson propagators. Indeed,
   we are dealing with a coupled set of Bethe Salpeter equations. 
  One way to deal with the Bethe Salpeter equation is to solve the integral 
  equation, in which case a form factor is usually put to make the integrals 
  converge \cite{Lahiff:1999ur,JimenezTejero:2009vq}. The cut off is fitted to
  some data, usually the mass of some state, or a scattering length. We note
  that in these and other works, the vector propagator is kept together
  with one additional form factor, usually a static form factor 
  \cite{JimenezTejero:2009vq}. Together they impose a range to the
  interaction. We would like to state here that if the form factor is going to
  be fitted to experiment, one can as well neglect the $\vec{q}$ dependence of 
  the propagator and fit a global form factor to the data. 
   Keeping this in mind, we resort to the easy, and equivalent
  procedure of not putting any form factor but cutting the loop integrals at a
  maximum value of the three-momentum. In other words, we would be introducing a
  global static form factor which is a theta function. The parameter to fit in
  this case if the cut off momentum, or taking into account the equivalence of
  this procedure with dimensional regularization \cite{ollerulf}, 
  fixing the subtraction
  constant. We  make use of this latter procedure. The approach has
   technical advantages; the potential now factorizes in the loops and then
   only the integrals of two propagators need to be performed and the set of
  Bethe Salpeter equations becomes now a set of algebraic equations. The
  procedure is equivalent to using the on-shell factorization, but with
  a potential with no $\vec{q}$ dependence which corresponds to a
  constant, $i.e.$ a local term with no left hand cut. This is the way our equations
  must be visualized and the precise meaning of the on-shell factorization,
  often employed when dealing with these equations.
  
  The arguments used here for the t-channel can be extended to the u-channel in
  the case of coupled channels, where one would bind the external particles and
  have real momenta. The corrections of these terms are then very small, as we
  saw, and of the type $(\vec{k}/M_V)^2$.
  
     There is another issue that must be raised at this point. As described in
 detail in  \cite{JimenezTejero:2009vq} one has some flexibility to describe the
 same results changing simultaneously the form factor and the strength of the
 interaction. We also observe the same pattern by changing simultaneously the
 potential and the cut off, or subtraction constant. This is important to keep
 in mind in order to correct for approximations made in the potential.
 In this sense, our
 procedure is as follows. We obtain the strength of the potential from the
 theoretical approach, then take a cut off around 1 GeV, typical of the
 effective theories in the range of energies that we are considering and see that
 we obtain a bound state or resonance close to some physical state, to which the
 identification is likely.  After that, there is a small tuning of the cut off
 or subtraction constant such as to get the position of this state at a precise
 experimental energy. Later on the same cut off, subtraction constant 
 in our case, is used 
 to determine all the amplitudes and bound states or  resonances in other
 channels.  Let us assume we had just one bound state not far from 
 threshold (the argument also holds
 qualitatively for many channels when one of them clearly dominates). Let us then keep 
 in mind that our description of
 the states comes through the couplings of the resonances to the different
 channels. In the case of just one channel the coupling is given in terms of the
 binding energy according to the Weinberg compositness condition
 \cite{weinberg,efimov,Baru:2003qq,Dong:2008gb}. The theorem can be restated in
 the formalism used here, as shown in \cite{Gamermann:2009fv}, where one can see
 that the coupling depends exclusively on the derivative of the $G$ function at
 the pole position independently of the potential. Our fitting procedure to 
 the mass of one
 state undoubtedly benefits from this theorem that guaranties a good coupling in
 the case of one channel, or dominance of one channel in the coupled channel
 case. The procedure of fixing one state also provides a better stability of 
 the theoretical results for the nearby
 states since the difference of masses can always be better predicted than
 absolute ones. 

     One more comment concerning the work of \cite{JimenezTejero:2009vq} is in
 order. The results "beyond the zero range approximation" reported there take
 into account the range of
 the exchanged vector mesons and an extra static form factor is used. 
 The main effect observed there is an increase of
 the non diagonal potentials mostly because of the energy transfer in the 
 vector propagators. This
 has as a result a moderate increase of the widths of the states. The positions
 are practically  unchanged. This energy transfer is sometimes non negligible
 in \cite{JimenezTejero:2009vq} because the authors  are dealing with 
 charmed mesons and baryons where the
 differences of masses with respect to non charmed hadrons are large. The same
 corrections are taken into account in \cite{MartinezTorres:2009xb}.  In the
 present case we
 deal with vector mesons in SU(3) which have very similar masses and that
 feature does not appear.
 
   After this discussion, justifying the  approach presented at the 
   beginning of section 2, we move on to show the results obtained.

\section{Results}

In this section we show results for the amplitudes obtained in the attractive
channels mentioned above. In figs.~\ref{fig:s0}, \ref{fig:sm1} 
and \ref{fig:sm2} we show the results for the channels
$S,I=0,1/2$; $0,3/2$; $-1,0$; $-1,1$; $-2,1/2$ and $-3,0$.  The value of $a_l(\mu)$
taken is -2.2 MeV.

\subsection{S,I=0,1/2 states}
\begin{figure}[htb]
\includegraphics[width=\textwidth]{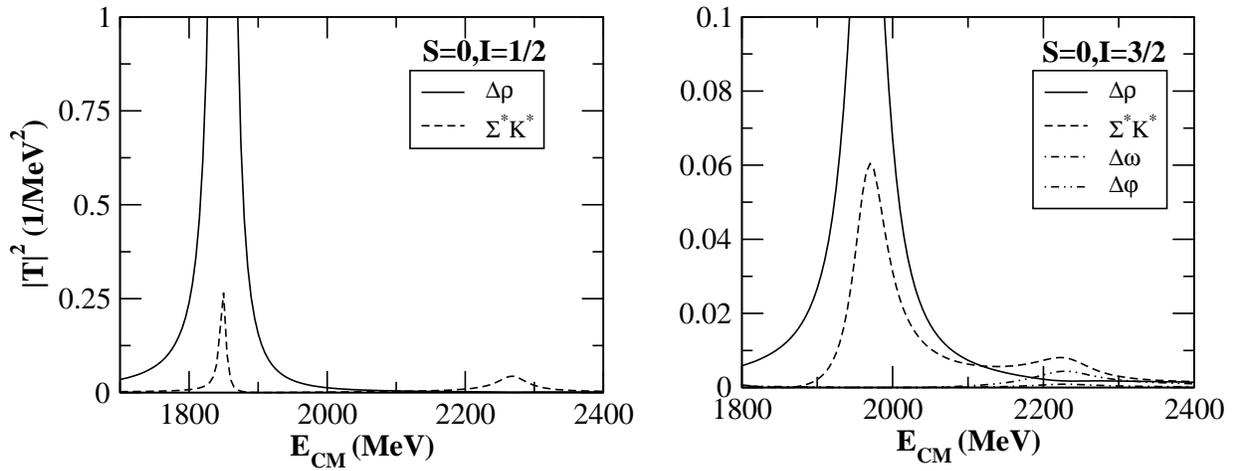}
\caption{$|T|^2$ for $S,I=0,1/2$ and $0,3/2$.}
\label{fig:s0}
\end{figure}

 The figures show peaks of $|T|^2$ which indicate the existence of a pole
 or a resonance. To further verify the guess, we look for poles in the complex
 plane as done in \cite{angels,ollerulf,Roca:2005nm}.
 The real part of the pole position provides the mass of the resonance and two
 times the imaginary part its width. The results can be seen in tables~\ref{tab:s0i12},
\ref{tab:s0i32}, \ref{tab:sm1i0}, \ref{tab:sm1i1}, \ref{tab:sm2i12}, \ref{tab:sm3i0}.
In fig.~\ref{fig:s0} we show the results for the $S,I=0,1/2$; $0,3/2$ channels. In the
$S,I=0,1/2$ sector (fig.~\ref{fig:s0}(left)) we see a clear peak around 1850 MeV, most pronounced
in the $\Delta \rho$ channel, but also visible in $\Sigma^* K^*$.
This is also reflected in the couplings in table~\ref{tab:s0i12}. This difference in the couplings and the large mass difference of the $\Sigma^* K^*$ state with
respect to the mass of the resonance makes the effect of this channel
essentially negligible and the state qualifies cleanly as a $\Delta \rho$
state, as was assumed in \cite{vijande}. The second weak structure around 2270
MeV on the $\Sigma^* K^*$ threshold does not correspond to a pole. Note,
however, that such a bump could be identified experimentally with a resonance.
In fact, there is a cataloged resonance $N^*(2200) (5/2^-)$ around that energy.

\subsection{S,I=0,3/2 states}

 In this case we see a clear peak in fig.~\ref{fig:s0}(right) that shows
 mostly in the $\Delta \rho$ channel. An
 inspection of table~\ref{tab:s0i32} also indicates substantial coupling to the
 $\Sigma^* K^*$ channel.
 This state was obtained in \cite{vijande} considering only the
$\Delta \rho$ channel. A posteriori, we corroborate the approach of \cite{vijande}
 which was justified there in the fact that the lowest mass states should be
 constructed with the lowest mass vector and meson. The channel $\Delta
 \omega$ which has about the same mass as that of the $\Delta \rho$ was
 ignored in \cite{vijande} since, as one can see in table~\ref{tab:cijs0i32}, it does not
 couple to the $\Delta \rho$ nor to itself.  Now the final coupling of the
$\Delta \omega$ to the resonance is not zero, although negligible. The reason
is that the coupled channel treatment allows this resonance to couple to
 $\Delta \omega$ through non diagonal transitions to the $\Sigma^* K^*$
state.  The weaker broad structure that appears around 2200 MeV in the
 $\Sigma^* K^*$ channel does not correspond to a pole in the complex plane.
 However, we should note that with a slightly weaker subtraction constant,
 $a=-2.1$, it shows much more neatly and around 2150 MeV, and could be
 associated with the cataloged $\Delta(2150)(1/2^-)$.

\subsection{S,I=-1,0 states}

\begin{figure}[htb]
\includegraphics[width=\textwidth]{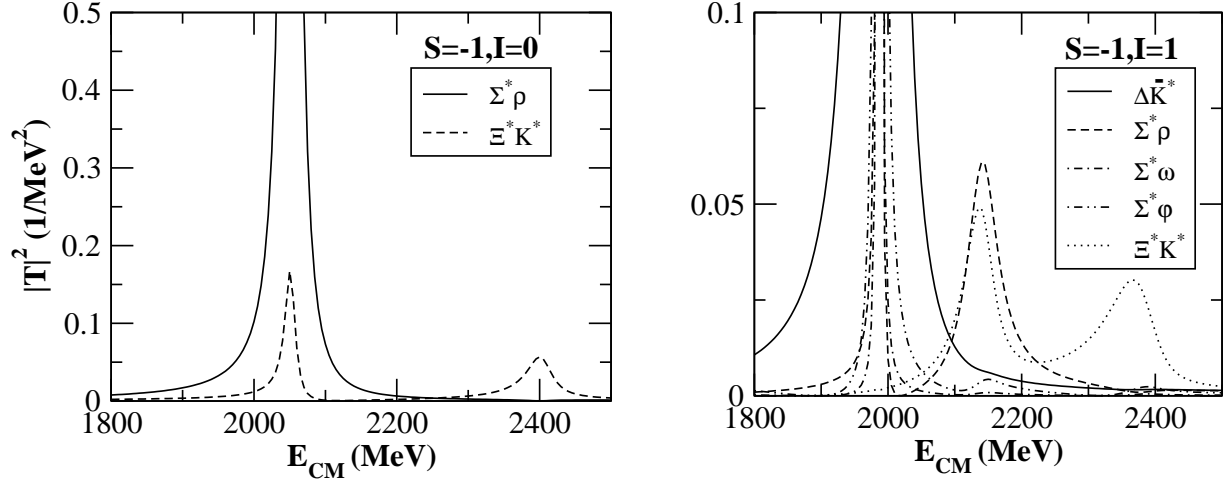}
\caption{$|T|^2$ for $S,I=-1,0$ and $-1,1$.}
\label{fig:sm1}
\end{figure}

In fig.~\ref{fig:sm1}(left) we find a clear peak around 2050 MeV which is most visible in the
$\Sigma^* \rho$ channel but also couples to the $\Xi^* K^*$. This is
corroborated by the strength of the couplings in table~\ref{tab:sm1i0}.
There is a weaker
structure in the $\Xi^* K^*$ channel around 2400 MeV (at threshold) which does not
correspond to a pole in the complex plane.

\subsection{S,I=-1,1 states}

Fig~\ref{fig:sm1}(right) shows this case where there are five coupled channels
producing three structures in the amplitude squared which indeed
correspond to three poles as shown in table~\ref{tab:sm1i1}. The first around 1990 MeV
couples strongly to $\Delta \bar{K}^*$, another one near 2150 MeV that couples strongly to
$\Sigma^* \rho$ and $\Xi ^* K^*$, and a third broader one around 2380 MeV that
couples mostly to $\Xi ^* K^*$.

\subsection{S,I=-2,1/2 states}

\begin{figure}[htb]
\includegraphics[width=\textwidth]{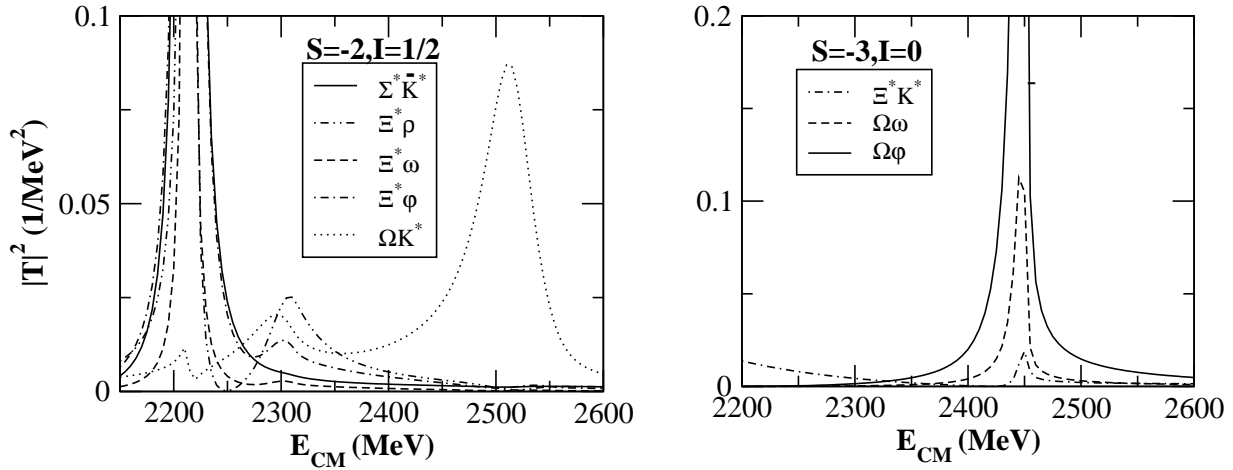}
\caption{$|T|^2$ for $S,I=-2,1/2$ and $-3,0$.}
\label{fig:sm2}
\end{figure}

In this case also there are five coupled channels. One finds three clear structures
in fig~\ref{fig:sm2}(left)  which correspond to three
poles of the scattering amplitude. The one around 2200 MeV
couples mostly to $\Sigma^* \bar{K}^*$ and $\Xi ^* \phi$
as seen from table~\ref{tab:sm2i12}. The second, around $2300$ MeV is seen
prominently in the $\Xi^*\rho$ channel.
The other
around 2520 MeV couples mostly to $\Omega K^*$ and appears as a
distinct peak structure in this channel.

\subsection{S,I=-3,0 states}

We see in fig.~\ref{fig:sm2}(right)  a
clear peak around 2450 MeV corresponding to a pole in the complex energy plane around
this energy. As
seen from table~\ref{tab:sm3i0},
this state couples mostly to $\Omega \phi$ and $\Omega \omega$.

\section{Comparison to data}

\begin{table*}[!ht]
      \renewcommand{\arraystretch}{1.5}
     \setlength{\tabcolsep}{0.2cm}
\begin{center}
\begin{tabular}{c|l|cc|lcclc}\hline\hline
$S,\,I$&\multicolumn{3}{c|}{Theory} & \multicolumn{5}{c}{PDG data}\\\hline
        & pole position &\multicolumn{2}{c|}{real axis} & name & $J^P$ & status & mass & width \\
        &               & mass & width & \\\hline
$0,1/2$ & $1850+i5$   & 1850  & 11  & $N(2090)$ & $1/2^-$ & $\star$ & 1880-2180 & 95-414\\
        &             &       &     & $N(2080)$ & $3/2^-$ & $\star\star$ & 1804-2081 & 180-450\\
        &       &  $2270(bump)$ &  & $N(2200)$ & $5/2^-$ & $\star\star$ & 1900-2228 & 130-400\\
\hline
$0,3/2$ & $1972+i49$  & 1971  & 52  & $\De(1900)$ & $1/2^-$ & $\star\star$ & 1850-1950 & 140-240 \\
    &             &       &     & $\De(1940)$ & $3/2^-$ & $\star$ & 1940-2057 & 198-460   \\
        &             &       &     & $\De(1930)$ & $5/2^-$ & $\star\star\star$ & 1900-2020  & 220-500   \\
    &             & $2200 (bump)$  &     & $\De(2150)$ & $1/2^-$ & $\star$ & 2050-2200  & 120-200  \\
\hline
$-1,0$  & $2052+i10$  & 2050  & 19  & $\Lambda(2000)$ & $?^?$ & $\star$  & 1935-2030 & 73-180\\
\hline
$-1,1$  & $1987+i1$   & 1985  & 10   & $\Sigma(1940)$ & $3/2^-$  & $\star\star\star$ &
1900-1950 & 150-300 \\
        & $2145+i58$  & 2144  & 57  & $\Sigma(2000)$ & $1/2^-$  & $\star$ & 1944-2004 &
    116-413\\
    & $2383+i73$  & 2370  & 99 & $\Sigma(2250)$ & $?^?$ & $\star\star\star$ & 2210-2280 &
    60-150\\
    &   &   &  & $\Sigma(2455)$ & $?^?$ & $\star\star$ & 2455$\pm$10 &
    100-140\\
\hline
$-2,1/2$ & $2214+i4$  & 2215  & 9  & $\Xi(2250)$ & $?^?$ & $\star\star$ & 2189-2295 & 30-130\\
     & $2305+i66$ & 2308  & 66 & $\Xi(2370)$ & $?^?$ & $\star\star$ & 2356-2392 & 75-80 \\
         & $2522+i38$ & 2512  & 60  & $\Xi(2500)$ & $?^?$ & $\star$ & 2430-2505 & 59-150\\
\hline
$-3,0$   & $2449+i7$   & 2445 & 13  & $\Omega(2470)$   & $?^?$   & $\star\star$ & 2474$\pm$12 & 72$\pm$33\\
 \hline\hline
    \end{tabular}
\caption{The properties of the 10 dynamically generated resonances and their possible PDG
counterparts. We also include the $N^*$ bump around 2270 MeV and the $\Delta^*$ bump around 2200 MeV. }
\label{tab:pdg}
\end{center}
\end{table*}

It is interesting to compare with data of the PDG. A summary is presented
in table~\ref{tab:pdg} in which the 10 dynamically generated states
have been listed along with their possible PDG counterparts including their present status and
properties. The two states
that we find for $S,I=0,1/2$ and $0,3/2$ around 1850 MeV and 1970 MeV respectively were
discussed in detail in \cite{vijande}. We do not discuss them further but
recall that there are indeed natural candidates for these states in the
$\Delta(1900)~ 3/2(1/2^-)$, $\Delta(1940)~ 3/2(3/2^-)$$\Delta(1930)~
3/2(5/2^-)$. As discussed there, some fine tuning of the subtraction constant
can make the masses agree better. But differences of 50 MeV or more are the
state of the art in determining masses in hadronic models. As to the $N^*$ states
around these energies, we mention, recalling \cite{vijande} that there is a
large dispersion of data in the masses but there are indeed several candidate
states like the $N^*(2090)$ $1/2(1/2^-)$,  $N^*(2080)$ $1/2(3/2^-)$ etc.
though some of the states predicted could be missing.

   The $\Lambda ^*$ state that we see in $S,I=-1,0$ around 2050 MeV could
   correspond to the $\Lambda(2000)$ with spin parity unknown. We do not see
   traces of the spin partners in this energy region.

 In the $\Sigma ^* $ sector we found three states around 1990, 2150, 2380 MeV.
 One can find two states around 2000 MeV in table~\ref{tab:pdg}, $\Sigma(1940)~(3/2^-)$  and
$\Sigma(2000)~(1/2^-)$. At higher energies there are several states. One of
them, the  $\Sigma^*(2250)$ classified with three stars, where some experiments
see two resonances, one of them with $5/2^-$. Another one is the  $\Sigma^*(2455) $
classified as bumps without spin and parity assignment. This region is the experimental
frontier in this sector and we hope that the findings of the present work
stimulate further work to complete the table. Indeed, the information we
provide, telling to which states the resonance couples most strongly could be
a guiding line for the search of these new states.

  In the $\Xi^*$, $S,I=-2,1/2$  sector we find three states around 2200 MeV,
  2300 MeV and 2520 MeV. As shown in table~\ref{tab:pdg} in the PDG,
  we find three states  $\Xi(2250)$, $\Xi(2370)$,
  $\Xi(2500)$ with spin and parity unknown. This sector is also poorly known but
  an experimental program is running at Jefferson Lab to widen the information
  in this sector \cite{Nefkens:2006bc,Price:2004xm}.

  Finally, in the $\Omega$ sector we find a clean state around 2450 MeV. Once
  again we find in the PDG the $\Omega(2470)$ without spin and parity
  assigned.

     The width of the resonances in the present framework is given by twice
     the imaginary part of the pole position in the complex energy plane. Also
     reported in the 'real-axis' column of table~\ref{tab:pdg} is the full width at half maximum obtained from
     the square of the amplitude plotted as a function of energy. However,
     we do not consider it meaningful to compare the widths of the dynamically generated states
     with those of the PDG in this work. Those reported here correspond only to
     decay into the vector-baryon decuplet states. In addition there are
     decays into a pseudoscalar meson and an octet baryon which can also be calculated
     in this approach, in a similar way as it was done in \cite{Molina:2008jw}
     to obtain the decay into $\pi \pi$ of the $\rho \rho$ states found there.
     Given the incomplete information about the states in the PDG in this
     area and the qualitative assignments that we can only do at present, its
     evaluation would not provide more help establishing the association of
     theoretical states with the  experimental ones, but when the situation
     improves, this work could be done to help understand better the spectrum of
     baryons.

\section{Conclusions}

We have studied the interaction of vector mesons with the decuplet of
baryons, looking for bound states or resonances, and we find ten  poles in
the scattering matrices of different strangeness and isospin channels. The
states are furthermore degenerate in $J^P=1/2^-, 3/2^-, 5/2^-$. We found
candidates in the PDG to be associated to the states found, but many of them
are missing. This should not be a surprise since the states we find are at the
frontier of the experimental research. The success of the method applied in
other sectors, as the meson sector, and even here for the lowest energy
states found, makes us confident that these states exist. The difficulties in
finding them are understandable. Most of the states reported in this energy
region are found in partial wave analysis of reactions using the available
pion, photon, electron or kaon beams. Some of these analyses require the
simultaneous consideration of many resonances, more than twenty in some cases,
which makes the conclusions problematic. Indeed, different solutions are
often found, concerning not only the properties but even the existence of some
resonances, depending on the type
of analysis or the assumptions made. New schemes become necessary to make
progress in this area. In this sense, the present theoretical work, as well as
others, where predictions are made regarding the channels to which the
resonances couple most strongly, could be considered as a guideline to search
for specific final states which could filter some of the resonances. We
envisage a fruitful new strategy of research along this line and encourage
future work in this direction.

\section*{Acknowledgments}
B. X. Sun would like to thank Li-Sheng Geng and R. Molina for useful discussions.
This work is partly supported by DGICYT contract number
FIS2006-03438.
This research is  part of the EU Integrated Infrastructure Initiative Hadron Physics Project
under  contract number RII3-CT-2004-506078. B. X. Sun acknowledges support from the National Natural Science Foundation of China under grant number 10775012.

\newpage

\section{Appendix}
\renewcommand{\arraystretch}{1.5}

\begin{table}[hbt]
\begin{center}
\begin{tabular}{c|cc}
\hline
 & $\De\rho$ & $\Sg K^{*}$ \\
\hline
$\De\rho$ & 5 & 2 \\
$\Sg K^{*}$ & & 2 \\
\hline
\end{tabular}
\caption{$C_{ij}$ coefficients for $S=0$, $I=\frac{1}{2}$.}
\label{tab:cijs0i12}
\end{center}
\end{table}

\begin{table}[hbt]
\begin{center}
\begin{tabular}{c|cccc}
\hline
 & $\De\rho$ & $\Sg K^{*}$ & $\De\omega$  &$\De\phi$\\
\hline
$\De\rho$   & 2 & $\sqrt{\frac{5}{2}}$ &0 & 0 \\
$\Sg K^{*}$ & & $-1$ & $\sqrt{\frac{3}{2}}$& $-\sqrt{3}$ \\
$\De\omega$ & & & 0  & 0 \\
$\De\phi$   & & & & 0 \\

\hline
\end{tabular}
\caption{$C_{ij}$ coefficients for $S=0$, $I=\frac{3}{2}$.}
\label{tab:cijs0i32}
\end{center}
\end{table}

\begin{table}[hbt]
\begin{center}
\begin{tabular}{c|cc}
\hline
 & $\Sg\rho$ & $\X K^{*}$ \\
\hline
$\Sg\rho$ & 4 & $\sqrt{6}$ \\
$\X K^{*}$ & & 3 \\
\hline
\end{tabular}
\caption{$C_{ij}$ coefficients for $S=-1$, $I=0$.}
\label{tabS-1I0}
\end{center}
\end{table}

\begin{table}[hbt]
\begin{center}
\begin{tabular}{c|ccccc}
\hline
 & $\De\ov K^{*}$ & $\Sg \rho$ & $\Sg\omega$ & $\Sg\phi$ & $\X K^{*}$\\
\hline
$\De\ov K^{*}$ & 4 & 1 & $\sqrt{2}$& -2 & 0 \\
$\Sg\rho$ & & 2 & 0 & 0 & 2 \\
$\Sg\omega$ & & &0& 0 & $\sqrt{2}$ \\
$\Sg\phi$   & & & & 0 & -2 \\
$\X K^{*}$ & & & & & 1 \\
\hline
\end{tabular}
\caption{$C_{ij}$ coefficients for $S=-1$, $I=1$.}
\label{tabS-1I1}
\end{center}
\end{table}

\begin{table}[hbt]
\begin{center}
\begin{tabular}{c|cc}
\hline
 & $\De\ov K^{*}$ & $\Sg \rho$ \\
\hline
$\De\ov K^{*}$ & 0 & $\sqrt{3}$ \\
$\Sg \rho$ & & $-2$ \\
\hline
\end{tabular}
\caption{$C_{ij}$ coefficients for $S=-1$, $I=2$.}
\label{tabS-1I2}
\end{center}
\end{table}

\begin{table}[hbt]
\begin{center}
\begin{tabular}{c|ccccc}
\hline
 & $\Sg\ov K^{*}$ & $\X \rho$ & $\X\omega$ & $\X\phi$& $\Om K^{*}$\\
\hline
$\Sg\ov K^{*}$ & 2 & 1 & $\sqrt{3}$  & $-\sqrt{6}$ & 0 \\
$\X\rho$ & & 2 & 0 & 0 & $\frac{3}{\sqrt{2}}$ \\
$\X\omega$ & & & 0 & 0 & $\sqrt{\frac{3}{2}vd_rev-0416-ch}$ \\
$\X\phi$   & & &   & 0 & $-\sqrt{3}$ \\
$\Om K^{*}$& & &   &   & 3 \\
\hline
\end{tabular}
\caption{$C_{ij}$ coefficients for $S=-2$, $I=\frac{1}{2}$.}
\label{tabS-2I1by2}
\end{center}
\end{table}

\begin{table}[hbt]
\begin{center}
\begin{tabular}{c|cc}
\hline
 & $\Sg\ov K^{*}$ & $\X \rho$ \\
\hline
$\Sg\ov K^{*}$ & $-1$ & 2 \\
$\X \rho$ & & $-1$ \\
\hline
\end{tabular}
\caption{$C_{ij}$ coefficients for $S=-2$, $I=\frac{3}{2}$.}
\label{tabS-2I3by2}
\end{center}
\end{table}

\begin{table}[hbt]
\begin{center}
\begin{tabular}{c|ccc}
\hline
 & $\X\ov K^{*}$ & $\Om \omega$ & $\Om \phi$ \\
\hline
$\X\ov K^{*}$ & 0 & $\sqrt{3}$ & $-\sqrt{6}$\\
$\Om \omega$ & & 0 & 0\\
$\Om \phi$ & & & 0 \\
\hline
\end{tabular}
\caption{$C_{ij}$ coefficients for $S=-3$, $I=0$.}
\label{tabS-3I0}
\end{center}
\end{table}

\begin{table}[hbt]
\begin{center}
\begin{tabular}{c|cc}
\hline
 & $\X\ov K^{*}$ & $\Om \rho$ \\
\hline
$\X\ov K^{*}$ & $-2$ & $\sqrt{3}$ \\
$\Om \rho$ & & $-2$ \\
\hline
\end{tabular}
\caption{$C_{ij}$ coefficients for $S=-3$, $I=1$.}
\label{tabS-3I1}
\end{center}
\end{table}

\begin{table}[hbt]
\begin{center}
\begin{tabular}{c|cc}
\hline
  $z_{R}$ & \multicolumn{2}{c}{$1850+ i5$} \\
\cline{2-3}
& $g_i$ & $|g_i|$ \\
\hline
$\De\rho$ & $4.9+i0.1$ & $4.9$  \\
$\Sg K^*$ & $1.7+i0.0$ & $1.7$  \\
\hline
\end{tabular}
\caption{The position of the pole and the coupling constant $g_{i}$ of the resonance for $S=0$, $I=\frac{1}{2}$ .}
\label{tab:s0i12}
\end{center}
\end{table}

\begin{table}[hbt]
\begin{center}
\begin{tabular}{c|cc}
\hline
  $z_{R}$ & \multicolumn{2}{c}{$1972+i49$} \\
\cline{2-3}
& $g_i$ & $|g_i|$  \\
\hline
$\De\rho$ & $5.0+i0.2$ & $5.0$ \\
$\Sg K^*$ & $3.9-i0.1$ & $3.9$  \\
$\De\omega$ & $-0.1+i0.2$ & $0.3$  \\
$\De\phi$ & $0.2-i0.4$ & $0.4$  \\
\hline
\end{tabular}
\caption{The position of the pole and the coupling constant $g_{i}$
of the resonance for $S=0$, $I=\frac{3}{2}$.}
\label{tab:s0i32}
\end{center}
\end{table}

\begin{table}[hbt]
\begin{center}
\begin{tabular}{c|cc}
\hline
  $z_{R}$ & \multicolumn{2}{c}{$2052+ i10$} \\
\cline{2-3}
& $g_i$ & $|g_i|$ \\
\hline
$\Sg\rho$ & $4.2+i0.1$ & $4.2$ \\
$\X K^*$ & $2.0+i0.1$ & $2.0$ \\
\hline
\end{tabular}
\caption{ The position of the pole and the coupling constant $g_{i}$ of the resonance for $S=-1$, $I=0$. }
\label{tab:sm1i0}
\end{center}
\end{table}

\begin{table}[hbt]
\begin{center}
\begin{tabular}{c|cc|cc|cc}
\hline
  $z_{R}$ & \multicolumn{2}{c|}{$1987 + i1$} &
\multicolumn{2}{c|}{$2145+ i58$} & \multicolumn{2}{c}{$2383+ i73$}  \\
\cline{2-7}
& $g_i$ & $|g_i|$ & $g_i$ & $|g_i|$ & $g_i$ & $|g_i|$ \\
\hline
$\De\ksb$ & $4.2+i0.0$ & $4.2$ & $0.7+i0.1$ & $0.7$ & $0.4+i0.4$ & $0.6$ \\
$\Sg \rho$ & $1.4+i0.0$ & $1.4$ & $-4.3-i0.7$ & $4.4$ & $0.4+i1.1$ & $1.2$ \\
$\Sg\omega$ & $1.4+i0.0$ & $1.4$ & $1.3-i0.4$ & $1.3$ & $-1.4-i0.4$ & $1.5$ \\
$\Sg\phi$ & $-2.1-i0.0$ & $2.1$ & $-1.9+i0.6$ & $2.0$ & $2.1+i0.6$ & $2.2$ \\
$\X K^*$ & $0.1-i0.0$ & $0.1$ & $-4.0-i0.1$ & $4.0$ & $-3.5+i1.5$ & $3.8$ \\
\hline
\end{tabular}
\caption{The position of the pole and the coupling constant $g_{i}$ of the resonance for $S=-1$, $I=1$.}
\label{tab:sm1i1}
\end{center}
\end{table}

\begin{table}[hbt]
\begin{center}
\begin{tabular}{c|cc|cc|cc}
\hline
  $z_{R}$ & \multicolumn{2}{c|}{$2214+i4$} & \multicolumn{2}{c|}{$2305+i66$}
   & \multicolumn{2}{c}{$2522 + i38$}  \\
\cline{2-7}
& $g_i$ & $|g_i|$ & $g_i$ & $|g_i|$  & $g_i$ & $|g_i|$ \\
\hline
$\Sg\ksb$ & $2.4+i0.1$ & $2.4$ & $0.8-i0.1$ & $0.8$ & $0.3+i0.3$ & $0.4$  \\
$\X \rho$ & $1.8-i0.1$ & $1.8$ & $-3.5-i1.7$ & $3.9$ & $0.2+i1.0$ & $1.0$  \\
$\X\omega$ & $1.7+i0.1$ & $1.7$ & $2.0-i0.7$ & $2.1$ & $-0.6-i0.3$ & $0.7$  \\
$\X\phi$ & $-2.5-i0.1$ & $2.5$ &  $-3.0+i1.0$ & $3.1$ & $0.9+i0.4$ & $1.0$  \\
$\Om K^*$ & $0.5-i0.1$ & $0.5$ & $-2.7-i0.8$ & $2.8$ & $-3.3+i0.9$ & $3.4$ \\
\hline
\end{tabular}
\caption{The position of the pole and the coupling constant $g_{i}$ of the resonance for $S=-2$, $I=1/2$.}
\label{tab:sm2i12}
\end{center}
\end{table}

\begin{table}[hbt]
\begin{center}
\begin{tabular}{c|cc}
\hline
  $z_{R}$ & \multicolumn{2}{c}{$2449 + i7$} \\
\cline{2-3}
& $g_i$ & $|g_i|$ \\
\hline
$\X\ksb$ & $1.0+i0.2$ & $1.0$ \\
$\Om\omega$ & $1.6-i0.2$ & $1.7$ \\
$\Om\phi$ & $-2.4+i0.3$ & $2.5$ \\
\hline
\end{tabular}
\caption{The position of the pole and the coupling constant $g_{i}$ of the resonance for $S=-3$, $I=0$.}
\label{tab:sm3i0}
\end{center}
\end{table}

\end{document}